\documentclass[english, prb, reprint, superscriptaddress]{revtex4-2}
\usepackage[utf8]{inputenc}    
\usepackage[T1]{fontenc} 
\usepackage{graphicx}
\usepackage{microtype} 
\usepackage{amsmath}  
\usepackage{amssymb} 
\usepackage{xcolor} 
\usepackage{ulem}
\usepackage{times,mathptmx}
\usepackage[colorlinks=true,citecolor=blue,linkcolor=blue]{hyperref}
\definecolor{RED}{rgb}{1,0,0}
\usepackage[english]{babel}
\DeclareUnicodeCharacter{00E9}{\'{e}}    
\DeclareUnicodeCharacter{00E1}{\'{a}}   
\DeclareUnicodeCharacter{00F3}{\'{o}}

\makeatletter

\renewcommand{\fnum@figure}{Fig.~\thefigure}

\makeatother

\usepackage{babel}

\begin{document}
\title{Symmetry-Breaking Induced Spin Transport and Magneto-Optical Regulation in 2D Altermagnet $\text{Ru}_2\text{MoSe}_4$}

\author{Wenpeng Wang}
\affiliation{College of Sciences, Northeastern University, Shenyang 110819, China}

\author{Hang Shi}
\affiliation{College of Sciences, Northeastern University, Shenyang 110819, China}

\author{Yuping Tian}
\affiliation{College of Sciences, Northeastern University, Shenyang 110819, China}

\author{Wei-Jiang Gong}
\address{College of Sciences, Northeastern University, Shenyang 110819, China}

\author{Xiangru Kong}
\email{kongxiangru@neu.edu.cn}
\affiliation{College of Sciences, Northeastern University, Shenyang 110819, China}

\begin{abstract}
Two-dimensional (2D) altermagnets (AMs) offer a compelling paradigm for advanced spintronics, yet their fully compensated macroscopic spin currents inherently limit practical device integration. In this work, using first-principles calculations and theoretical analysis, we demonstrate that the 2D material $\text{Ru}_2\text{MoSe}_4$ hosts AM ground state protected by $\text{S}_{4z}\bm{T}$ symmetry. Using uniaxial strain modulation and stacking configuration, we show that the monolayer and the AC-stacking bilayer $\text{Ru}_2\text{MoSe}_4$ host fully spin-polarized currents, piezomagnetically induced net magnetization, and the magneto-optical Kerr effect. Our findings establish $\text{Ru}_2\text{MoSe}_4$ as a tunable platform, offering a feasible mechanism to simultaneously trigger electrical spin transport signals and amplify optical readout signatures for next-generation spintronics and valleytronics.

\end{abstract}

\maketitle
\section{Introduction}
As a new class of magnetic materials that break time-reversal symmetry, altermagnets (AMs) integrate the merits of both ferromagnetic (FM) and antiferromagnetic (AFM) materials, providing a novel paradigm for advanced spintronics ~\cite{baltz_antiferromagnetic_2018,doi:10.1126/sciadv.aaz8809,PhysRevLett.130.216702,PhysRevLett.126.127701,doi:10.1126/sciadv.adj4883,guo_valley_2025,PhysRevLett.133.056401}. In reciprocal space, AMs feature a spin-split band structure, despite maintaining an absolute absence of macroscopic net magnetization in real space ~\cite{10.1063/5.0312073,PhysRevLett.132.236701,krempasky_altermagnetic_2024,PhysRevLett.133.086503,PhysRevLett.130.216701,jiang_strain-engineering_2025,reimers_direct_2024,r8nc-dpt8}. The alternating spin configuration inherent in AMs serves as the origin of a range of unconventional transport responses. Notable examples include the anomalous Hall effect (AHE) ~\cite{feng_anomalous_2022,doi:10.1126/sciadv.aaz8809,PhysRevB.111.184408,PhysRevLett.130.036702,PhysRevB.111.184407,PhysRevLett.133.086503} and the topological transport in quantum spin Hall (QSH) systems ~\cite{zbbr-426l,chen_quantum_2025,zhang_quantized_2025}.
The emergence of two-dimensional (2D) AMs, particularly in bilayer van der Waals systems ~\cite{PhysRevB.110.014442,PhysRevLett.133.166701,j4gp-ctxj,jiang_controlling_2018}, introduces the layer degree of freedom as a powerful new control parameter for manipulating the intrinsic non‑relativistic spin splitting of AMs themselves ~\cite{zeng_bilayer_2024,doi:10.1021/acs.nanolett.4c06037,tsymbal_two-dimensional_2021}.

Despite these advantages, the macroscopic spin currents in pristine AMs are often fully compensated due to the protection of strict magnetic spin-group and crystal symmetries ~\cite{song_observation_2026}. Consequently, although microscopic spin splitting of band structure is present, ideal lattices fail to generate a net spin current or macroscopic magnetization, which is a significant bottleneck for signal output in practical devices ~\cite{PhysRevLett.126.127701,zhang_crystal-symmetry-paired_2025}. Moreover, the high-efficiency, non-destructive detection of net spin currents remains a formidable challenge. Traditional electrical contact probing is often plagued by interface scattering ~\cite{doi:10.1126/sciadv.adj4883,PhysRevB.109.115102,zeng_observation_2024,jeong2026altermagnetic,PhysRevLett.130.036702}. As a superior non-contact readout mechanism, the magneto-optical Kerr effect (MOKE) is particularly useful for detecting, measuring, and manipulating magnetic systems ~\cite{doi:10.1021/acs.nanolett.3c05052,higo_large_2018,balk_comparing_2019,huang_layer-dependent_2017}, including 2D magnets. Therefore, there is an urgent need for a mechanism that can simultaneously trigger transport signals and amplify optical signatures.

Traditionally, the physical origin of MOKE is ascribed to the synergy between spin-orbit coupling (SOC) and time-reversal symmetry breaking ~\cite{10.1088/0034-4885/59/12/003,doi:10.1021/acs.nanolett.3c05052,oppeneer_ab_1992}. However, in AMs, spin-momentum locking can induce MOKE even in the complete absence of SOC ~\cite{liu_uncompensated_2025}. In first-principles calculations, the dynamic conductivity $\sigma_{\alpha\beta}(\omega)$ is calculated using the Kubo-Greenwood formula ~\cite{haider2017review}:
\begin{equation}
\sigma_{\alpha\beta}(\omega) = \frac{ie^2\hbar}{V} \sum_{\mathbf{k}, n \neq m} \frac{f_{n\mathbf{k}} - f_{m\mathbf{k}}}{\epsilon_{m\mathbf{k}} - \epsilon_{n\mathbf{k}}} \frac{\langle \psi_{n\mathbf{k}} | \hat{v}_\alpha | \psi_{m\mathbf{k}} \rangle \langle \psi_{m\mathbf{k}} | \hat{v}_\beta | \psi_{n\mathbf{k}} \rangle}{\epsilon_{m\mathbf{k}} - \epsilon_{n\mathbf{k}} - \hbar\omega - i\eta},
\end{equation}
where sums over all wave vectors \(k\) in the Brillouin zone and all distinct band pairs \((n,m)\). Each term represents the contribution from an interband transition of an electron from an occupied state \(n\) to an unoccupied state \(m\), where \(\hbar \omega\) is the photon energy and the small parameter \(\eta\) is the energy broadening (carrier lifetime effect). As a non-contact and highly sensitive technique, MOKE probe the spin polarization and magnetically ordered structures of electrons through changes in the polarization of reflected light, which serves as a cornerstone methodology in magnetism research, with its recent extension to FM ~\cite{yang_first-_2022}, AFM ~\cite{sivadas_gate-controllable_2016,wang_electric_2023,PMID:38525906} and AMs ~\cite{q8ym-l2zt,10.1088/0256-307X/43/2/020713,Liu2025UncompensatedLD,gray_time-resolved_2024} systems emerging as a prominent frontier.

The other routine for characterizing magnetic materials can use the Boltzmann transport equation in the relaxation time approximation ~\cite{surgers_anomalous_2016}, which calculate the electrical conductivity \(\sigma\) as a conductivity tensor ~\cite{RevModPhys.82.1539,smejkal_crystal_2020}.
The semiclassical Boltzmann transport theory describes the phase-space evolution of the carrier distribution function under non-equilibrium conditions, as well as the resulting macroscopic charge and energy transport processes ~\cite{madsen_boltztrap_2006}. In first-principles calculations, the direct-current (DC) electrical conductivity tensor is conventionally evaluated within the relaxation time approximation via the following formulation ~\cite{madsen_boltztrap_2006}:
\begin{equation}
\sigma_{\alpha\beta}(\mu, T) = e^2 \int d\epsilon \left[-\frac{\partial f}{\partial \epsilon}\right] \Sigma_{\alpha\beta}(\epsilon),
\end{equation}
where the transport distribution function \(\Sigma_{\alpha\beta}(\epsilon) = \frac{\tau}{V} \sum_{n,\mathbf{k}} v_{n\mathbf{k},\alpha} v_{n\mathbf{k},\beta} \, \delta(\epsilon - \epsilon_{n\mathbf{k}})\) relies solely on intra-band group velocities $v_{n\mathbf{k}} = \hbar^{-1} \nabla_{\mathbf{k}} \epsilon_{n\mathbf{k}}$. 
By enabling a direct comparison between theoretical computations and experimental measurements, this theory provides a robust platform for validating physical models and exploring novel physical phenomena ~\cite{PhysRevResearch.6.043185}.

In our work, using first-principles calculations and theoretical analysis, we propose that $\text{Ru}_2\text{MoSe}_4$ hosts AM ground state. Notably, two orthogonal symmetry-breaking strategies, which contain stacking configuration modulation and in-plane uniaxial strain, can break the compensatory symmetry of the lattice, simultaneously triggering a piezomagnetically induced net magnetization and a controllable macroscopic net spin current. We further show that the AC-stacked bilayer exhibits dramatically enhanced symmetry-breaking responses compared to the monolayer $\text{Ru}_2\text{MoSe}_4$, with the interlayer coupling boosting the MOKE signal by an order of magnitude. Our results provide a versatile platform for electrical-optically controllable AMs and next-generation spintronic applications.

\section{COMPUTATIONAL details}
First-principles calculations were performed within the framework of density functional theory (DFT) using the Vienna \(ab initio\) Simulation Package (VASP) ~\cite{PhysRev.140.A1133,PhysRevB.54.11169,kresse_efficiency_1996}. The strong correlation effects of Ru-4\(d\) electrons were addressed using the rotationally invariant PBE+U scheme for structural optimization and electronic structure calculations, with an effective Hubbard correction applied on-site of \(\mathrm{U_{eff} = 4.0}\) \(\mathrm{eV}\). Besides, the band structure with \(\mathrm{U_{eff} = 2.0, 3.0}\) and 5.0 \(\mathrm{eV}\) were shown in Fig.~S10 in Supplemental Material (SM). A cut-off energy of 500 \(\mathrm{eV}\) was employed for structural relaxation, with ionic optimization continuing until the Hellmann-Feynman forces in each atom were reduced below 0.01 \(\mathrm{eV}/\mathring{A}\), and the criterion of convergence of the energy was set to \(10^{-8}\) \(\mathrm{eV}\). Integration over the Brillouin zone (BZ) utilized a \(\Gamma\)-centered \(12\times12\times1\) Monkhorst-Pack k-point mesh ~\cite{PhysRevB.13.5188}, while the Berry curvature integration employed a significantly denser \(801\times801\times1\) grid. A large vacuum layer of 25 \(\mathring{A}\) was applied along the z/c-direction to eliminate artificial interactions between adjacent periodic images. Phonon spectra was computed using a \(3\times3\times1\) supercell within the framework of density functional perturbation theory (DFPT), as implemented in the PHONOPY code ~\cite{togo_first_2015}. The chemical bonding analysis was performed using the projected crystal orbital Hamilton population (pCOHP) method, as implemented in the LOBSTER program ~\cite{doi:10.1021/jp202489s,https://doi.org/10.1002/jcc.24300}. The input wavefunctions were obtained from static DFT calculations using VASP with the projector-augmented wave (PAW) method ~\cite{PhysRevB.50.17953,PhysRevB.59.1758}. 
The tight-binding Hamiltonian (TBH) was derived using maximally localized Wannier functions (MLWFs), constructed via the Wannier90 package ~\cite{mostofi_updated_2014}. The MLWFs are constructed by projecting Ru-\(d\), Mo-\(d\) and Se-\(p\) orbitals, thereby capturing the fundamental low-energy electronic behavior.
Based on the TBH, the Boltzmann transport properties were calculated using the Boltzmann transport equation ~\cite{pizzi_boltzwann_2014}. 
The Boltzmann transport coefficients were evaluated at a fixed temperature of 300 K, with a constant relaxation time of \(1\times{10}^{-14}\) \(s\). The circular dichroism was calculated using the VASPBERRY code ~\cite{Ponce_VASPBERRY_2021}, which postprocesses the DFT wavefunctions obtained from a static VASP calculation.

\section{RESULTS AND DISCUSSION}
\subsection{Structural Stability and Magnetic Symmetry}
\begin{figure}[t]
    \centering
    \includegraphics[width=8.5cm]{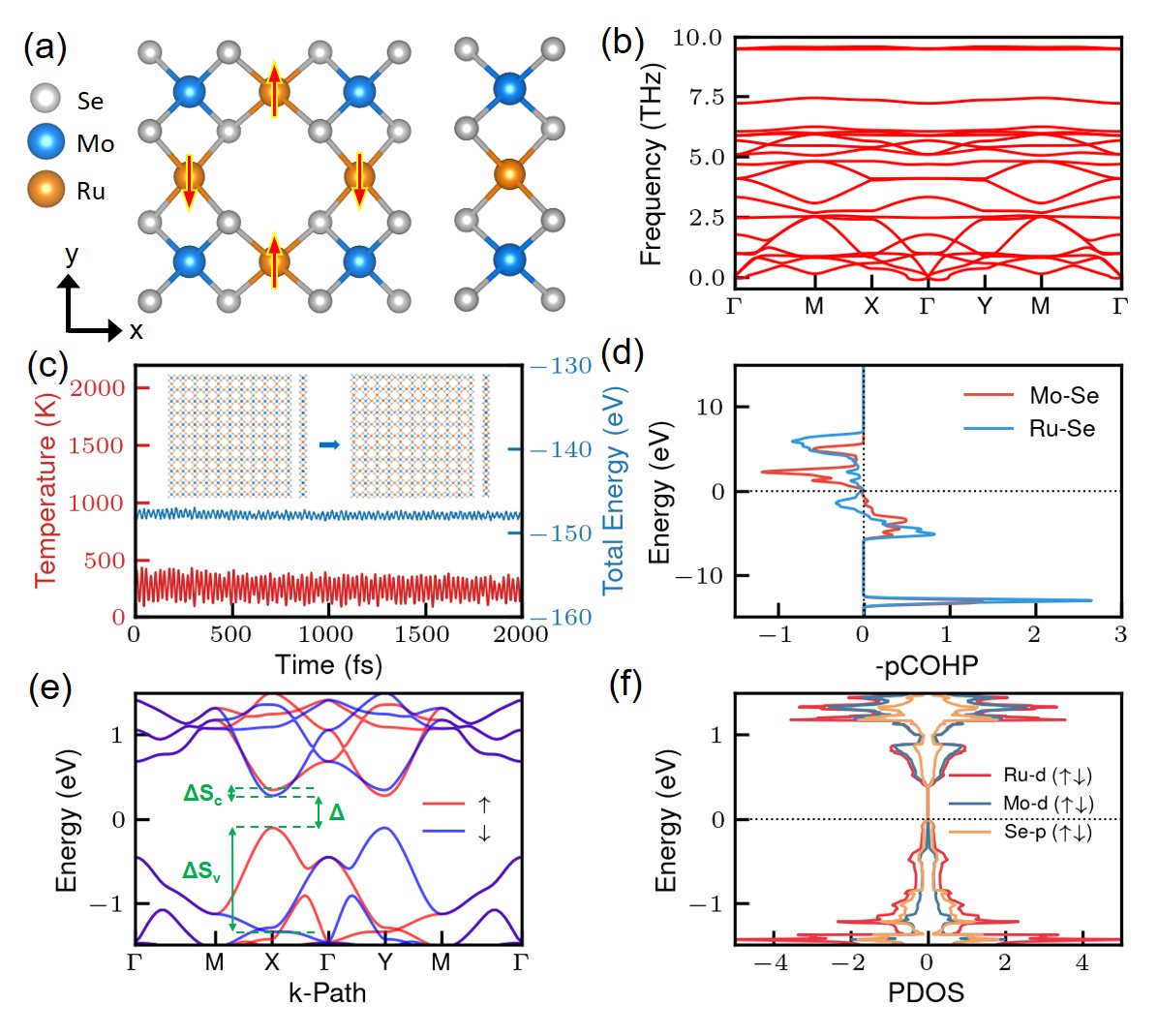}
    \caption{(a) Atomic structure of monolayer $\text{Ru}_2\text{MoSe}_4$, where arrows indicate the orientation of the magnetic moments applied to the Ru atoms. (b) Phonon dispersion spectrum for monolayer $\text{Ru}_2\text{MoSe}_4$. (c) The results of AIMD simulation for monolayer $\text{Ru}_2\text{MoSe}_4$ at 300 K. The inset illustrates the structural evolution by comparing the configurations before and after the simulation. (d) Calculated pCOHP for monolayer $\text{Ru}_2\text{MoSe}_4$. Positive and negative values represent bonding and antibonding states, respectively. (e) The electronic band structure without considering SOC. Here, $\Delta S_{c}$ denotes the difference in conduction band spin-splitting between the X and Y valleys, $\Delta S_{v}$ represents the valence band spin-splitting between the X and Y valleys, and $\Delta$ corresponds to the energy gap between the conduction and valence bands at that valley. (f) The PDOS for monolayer $\text{Ru}_2\text{MoSe}_4$. Positive and negative values on the horizontal axis correspond to the spin-up and spin-down channels, respectively.}
    \label{fig:1}
\end{figure}

$\text{Ru}_2\text{MoSe}_4$ monolayer has a square lattice with seven atoms in one primitive cell, belonging to the \(\text{P}\overline{4}2\text{m}\) space group (No.111) with \(\text{D}_{2d}\) point group, which possesses \({\text{E}, \text{C}_{2z}, \text{S}_{4z}, \text{C}_{2y}}\) crystal generators ~\cite{10.1524/zkri.2006.221.1.15}. When considering the collinear AFM order with Ru magnetic moments of \(\pm5\) \(\mu_B\) along the z axis, the magnetic symmetry is described by the magnetic space group \(\text{P}\overline{4}'2'\text{m}\) (No. 111.253). It has a sandwich structure that the Ru and Mo atomic sheet lies between two Se sheets in a symmetric fashion as the building block shown in Fig.~\ref{fig:1}(a). Fig.~\ref{fig:1}(b) shows that there is no obvious imaginary frequency, indicating that the $\text{Ru}_2\text{MoSe}_4$ monolayer is dynamically stable. The \(ab\) \(initio\) molecular dynamics (AIMD) simulation of the $\text{Ru}_2\text{MoSe}_4$ monolayer shown in Fig.~\ref{fig:1}(c) reveals that the nanostructure is almost unchanged under 300K. $\text{Ru}_2\text{MoSe}_4$ monolayer has excellent thermal stability at the room temperature due to the very small fluctuations of both total energy and temperature manifest with respect to the elapsed time.

According to the magnetic point group analysis, the magnetic structure of monolayer $\text{Ru}_2\text{MoSe}_4$ possesses $\text{S}_{4z}\bm{T}$ symmetry. Crucially, in conventional AFM, sublattices with opposite spin are typically related by inversion ($\bm{I}$) or translation ($\tau$) operation. However, in $\text{Ru}_2\text{MoSe}_4$, the sublattices are connected via the $\text{S}_{4z}$ roto-inversion operation. This absence of $\bm{PT}$ or $\tau\bm{T}$ symmetry, combined with the presence of a rotation-type operation connecting magnetic sublattices, classifies the system as an AM. 
Besides, from an electronic perspective, the $\text{S}_{4z}$ symmetry operation transforms the wave vector coordinates as $(k_x, k_y) \to (k_y, -k_x)$, thereby protecting the energetic degeneracy between the $X$ and $Y$ valleys in the reciprocal space while allowing for momentum-dependent spin splitting.

To ensure the magnetic ground state of the $\text{Ru}_2\text{MoSe}_4$ monolayer, FM and AFM configurations are considered, which contain four different magnetic orders, as shown in Fig.~S1. The calculated energy differences between the FM configuration and the AFM1-AFM3 configurations summarized in Table~S1 in SM, which can be defined as \(\Delta \text{E} = \text{E}_{\text{FM/AFM2/AFM3}} - \text{E}_{\text{AFM1}}\). This indicates the AM state is the ground state, which are shown in Fig.~\ref{fig:1}(a). The directional dependence of MAE, presented in Fig.~S2, indicates that the $\text{Ru}_2\text{MoSe}_4$ monolayer possesses an in-plane easy axis along the $x$ direction. 

To construct the bilayer structure of stacked monolayer $\text{Ru}_2\text{MoSe}_4$, we considered four distinct stacking configurations shown in Fig.~S4: the direct stacking, sliding along the \(x\) axis, \(y\) axis and their diagonal by half a lattice constant, which are named as AA, AB, AD and AC, respectively. The structural stability of these configurations was systematically evaluated, with the results presented in Fig.~S5 in SM. Among them, the AB and AD stackings exhibit pronounced imaginary phonon frequencies, indicating dynamical instability, whereas no imaginary frequencies are observed for the AA and AC configurations, confirming their dynamical stability.

Subsequently, we calculated the ground-state energies of the AA and AC stacking structures and found that the AC stacking possesses a lower ground-state energy than that in the AA configuration, which was shown in Table.~S2, suggesting that the AC structure is the most energetically favorable bilayer $\text{Ru}_2\text{MoSe}_4$. Given that the AC stacking demonstrates the highest thermodynamic stability, our subsequent investigation of physical properties focuses primarily on the monolayer and the ground-state AC bilayer structure.


To ensure if the AC-bilayer $\text{Ru}_2\text{MoSe}_4$ sustains alternating magnetic order, we performed a comparative magnetic symmetry analysis for bilayers of $\text{Ru}_2\text{MoSe}_4$. 
Upon stacking into the AC configuration, the symmetry evolves into the $\text{P}\overline{4}'2'_1\text{m}$ group (No. 113.269). 
The diagonal translation breaks the original mirror symmetry but introduces glide-like magnetic operations. Specifically, the magnetic group $\text{P}\overline{4}'2'_1\text{m}$ contains operations that combine a spatial transformation with time-reversal ($\bm{T}$), such as $\{2_x | 1/2, 1/2, 0\}\bm{T}$. These operations effectively map the magnetic moments of the upper layer to those of the lower layer with an associated spin flip ($M_z \to -M_z$). Consequently, the AC-stacking $\text{Ru}_2\text{MoSe}_4$ is fully symmetry-compatible with the AM state. In what follows, we perform first-principles calculations and a detailed analysis of the electronic band structures for these distinct systems.

\subsection{Electronic Structure and Spin-Valley Coupling}
The electronic properties and magnetic nature of the monolayer $\text{Ru}_2\text{MoSe}_4$ are further elucidated by investigating its spin-resolved band structure and projected density of states (PDOS), as illustrated in Fig.~\ref{fig:1}(e) and (f). The results obtained without considering SOC confirm that $\text{Ru}_2\text{MoSe}_4$ is a semiconductor featuring a direct band gap of 381 meV situated at the X and Y valleys. The PDOS reveals that the electronic states in the vicinity of the Fermi level ($E_F$) are predominantly governed by the Ru-$d$ orbitals, with both the valence band maximum (VBM) and conduction band minimum (CBM) exhibiting a high density of ruthenium states. 

The nature of AM system is manifested by the distinct energy shift between spin-up and spin-down channels in the band structure, leading to a momentum-dependent spin splitting in the band structure. The magnitude of this splitting, defined as $\Delta S_{c/v} = E^{\text{up}}_{c/v}(\mathbf{k}) - E^{\text{down}}_{c/v}(\mathbf{k})$, reaches 66.8 meV and 1188.5 meV for the conduction and valence bands at the $X$ valley, respectively. Notably, $\Delta S_{c/v}$ exhibits an antisymmetric distribution under mirror reflection in momentum space and $\Delta S_v$ is remarkably larger than that observed in other AMs such as $\text{Co}_2\text{MoSe}_4$~\cite{zhang_multiple_2025}. This pronounced splitting arises from the breaking of the combined space-inversion, time-reversal, and translation symmetries, while remaining protected by the $S_{4z}\bm{T}$ magnetic symmetry. The non-overlapping spin-resolved energy bands and the inherent coupling between X/Y valleys and up/down spin provide a robust mechanism for valley-specific band engineering. Consequently, monolayer $\text{Ru}_2\text{MoSe}_4$ emerges as a promising candidate for integrating properties of AMs into next-generation spintronic and valleytronic applications. Besides, in Fig.~\ref{fig:1}(d), the pCOHP analysis demonstrates robust bonding features for both Mo-Se and Ru-Se pairs. Notably, Ru-Se exhibits partially filled antibonding states in the vicinity of the Fermi level, in contrast to the virtually unoccupied antibonding orbitals of Mo-Se. Such electronic variations imply that while Mo-Se interactions are fundamental to structural integrity, the marginal filling of antibonding states in Ru-Se likely facilitates higher performance in electrochemical catalysis or charge transport.

\begin{figure}[t]
    \centering
    \includegraphics[width=8cm]{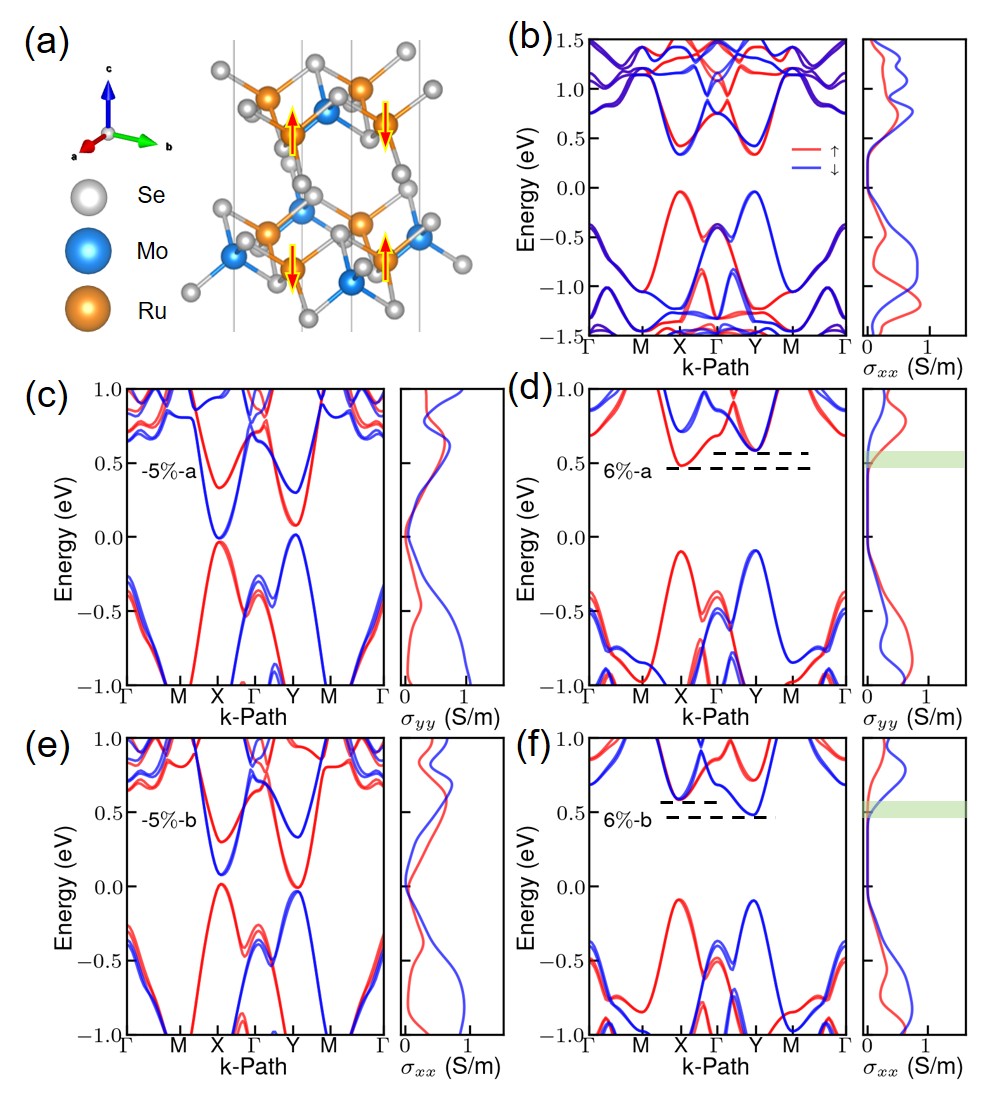}
    \caption{(a) Atomic structure of bilayer $\text{Ru}_2\text{MoSe}_4$ in the AC stacking configuration, where arrows indicate the orientation of the magnetic moments on the Ru atoms. (b) (Left) Electronic band structure of AC-stacked bilayer $\text{Ru}_2\text{MoSe}_4$ in the absence of SOC and external strain. (Right) Corresponding spin-resolved $\sigma_{xx}$. (c–f) Spin-resolved electronic band structures under uniaxial strain applied along the $a$-direction and $b$-direction. (c, e) is the uniaxial compressive (\(5\%\)) strain and (d, f) is the uniaxial tensile (\(6\%\)) strain. The dashed lines in (d) and (f) denote the positions of valley polarization. In the results of $\sigma_{xx}$ on the right, the green shaded regions highlight the areas where a net spin current is generated.}
    \label{fig:5}
\end{figure}

Upon assembling two $\text{Ru}_2\text{MoSe}_4$ monolayers into an AC-stacked bilayer, the emerging interlayer interactions modulate the electronic band structure.
In Fig.~\ref{fig:5}(b), a significant band gap of $461~\text{meV}$ is clearly discernible at the X and Y high-symmetry points, which is larger than in monolayer $\text{Ru}_2\text{MoSe}_4$ due to the strong interlayer orbital hybridization. The calculated PDOS (Fig.~S7(c)) indicates that the electronic characteristics near the Fermi level ($E_F$) are primarily governed by the Ru $d$-orbitals, which is similar to monolayer $\text{Ru}_2\text{MoSe}_4$.

Besides, it is important to note that both the monolayer and bilayer systems feature spin-polarized energy bands that manifest robust spin-valley coupling. Notably, the conduction band states at the X and Y valleys are explicitly locked to spin-down and spin-up configurations, respectively. Such strong spin-valley locking affords an effective mechanism for the synergistic manipulation of spin and valley degrees of freedom. Although spin splitting emerges in the system at this stage, the net spin polarization remains fully compensated.
This microscopically spin-polarized band structure provides the foundation for macroscopic spin transport. To evaluate its potential for spintronic applications, we employ the semiclassical Boltzmann transport theory to investigate the intrinsic transport properties of the system in the absence of external perturbations.

\subsection{Intrinsic Transport and Spin Current Detection Device}
\begin{figure}[t]
    \centering
    \includegraphics[width=8cm]{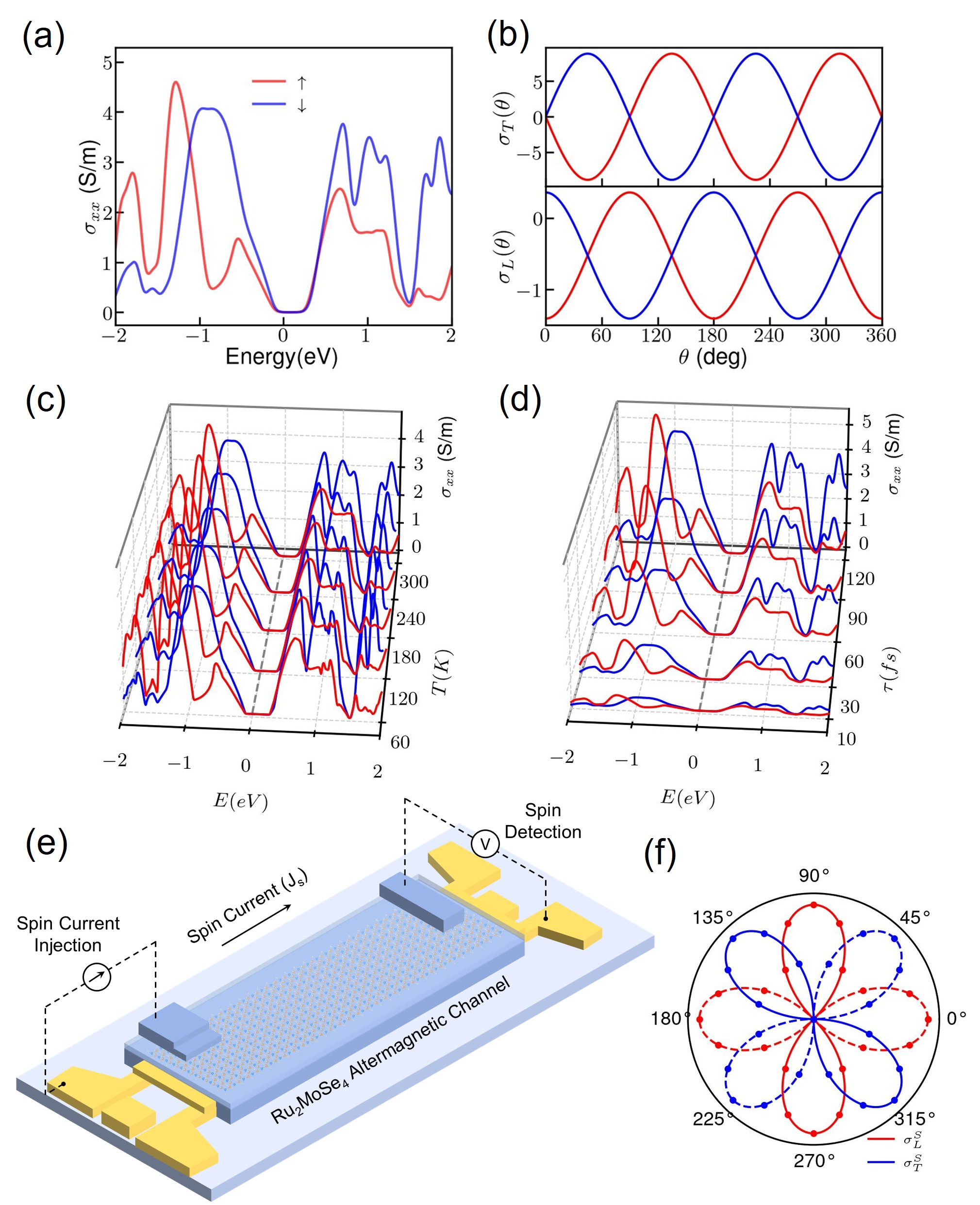}
    \caption{Transport properties of monolayer $\text{Ru}_2\text{MoSe}_4$. (a) Spin-resolved longitudinal conductivity $\sigma_{xx}$, which is a function of energy. (b) Angular dependence of the longitudinal ($L$) and transverse ($T$) charge conductivities as a function of the electric field direction $\theta$. (c, d) Spin-resolved $\sigma_{xx}$ calculated at various (c) temperatures and (d) electronic relaxation times $\tau$. (e) Schematic illustration of the experimental setup for spin current detection. (f) Corresponding angular dependence of the longitudinal and transverse spin conductivities.}
    \label{fig:4}
\end{figure}

We calculated the spin-resolved charge conductivity ($\sigma_{xx}$) of the system based on Boltzmann transport theory with a constant-relaxation-time approximation. 
As shown in Fig.~\ref{fig:4}(a), which shows $\sigma_{xx}$ as a function of energy relative to the Fermi level, the conductivity remains near zero around the Fermi level, signifying that the material is insulating near the Fermi level, consistent with semiconducting behavior. At higher or lower energies, non-zero conductivities are observed with distinct peaks for spin-up (red line) and spin-down (blue line) states, showcase that unequal conductivities of spin-up and spin-down states are induced by the anisotropic band spin splitting. Moreover, we study the angular dependence of longitudinal ($\sigma_L^{\uparrow,\downarrow}$) and transverse ($\sigma_T^{\uparrow,\downarrow}$) charge conductivities. Varying with electric field direction $\theta$ at a constant energy, as presented in Fig.~\ref{fig:4}(b), both $\sigma_T^{\uparrow,\downarrow}$ and $\sigma_L^{\uparrow,\downarrow}$ conductivities change with $\theta$ with a robust period of $180^\circ$. 
Furthermore, the net spin conductivities, which are given by $\sigma_L^S = \sigma_L^\uparrow - \sigma_L^\downarrow = -2 \sigma_0 cos2\theta$ and $\sigma_T^S = \sigma_T^\uparrow - \sigma_T^\downarrow = -2 \sigma_0 sin2\theta$ ~\cite{wu_valley-related_2024}, are also calculated, as drawn in Fig.~\ref{fig:4}(f).
The angular dependence of $\sigma_L^S$ forms lobes aligned with the crystal axes ($0^\circ$, $90^\circ$, $180^\circ$, $270^\circ$), showcasing peaks in longitudinal net spin generation along these directions. Conversely, the transverse net spin conductivity $\sigma_T^S$ displays a $45^\circ$ phase shift than $\sigma_L^S$. At $45^\circ$, vanishing $\sigma_L^S$ and maximum $\sigma_T^S$ give rise to a pure transverse current.


Besides, the impact of the electronic temperature ($T$) and relaxation time ($\tau$) on the $\sigma_{xx}$ spectra is also studied. As shown in Fig.~\ref{fig:4}(c), at temperatures ranging from 60 K to 300 K, the calculated $\sigma_{xx}$ spectra remain virtually identical, demonstrate robust transport properties against thermal effects in this regime. Conversely, as depicted in Fig.~\ref{fig:4}(d), $\sigma_{xx}$ is shown to scale linearly with the electronic relaxation time varying from 10 fs to 120 fs. However, the relative size, spin-resolved characteristics, and overall anisotropic features between spin-up and spin-down states remain basic similar for different electronic relaxation times. Therefore, the spin current generation characteristics of $\text{Ru}_2\text{MoSe}_4$ are fundamentally robust to electronic temperature and relaxation time, suggest potent applications in spintronic devices, possibly exhibit long spin depolarization times in the absence of SOC.

Based on transport analysis, it is crucial to further establish their experimental accessibility and device relevance. Thus, we propose a feasible measurement scheme for detecting the generated spin currents, as illustrated in Fig.~\ref{fig:4}(e). The AM layer (central channel) is contacted by source and drain electrodes that apply an in-plane electric field E along the \(x\)-direction. Owing to the anisotropic band spin splitting characteristic of the AM state, spin-up and spin-down electrons carry equal but opposite transverse conductivities (\(\sigma_T^\uparrow=-\sigma_T^\downarrow\)), while longitudinal charge conduction remains spin-polarized. Under the applied electric field, a longitudinal charge current flows collinear with E, whereas the transverse charge currents from the two spin channels cancel exactly, resulting in a net transverse pure spin current \(J_S=J_\uparrow-J_\downarrow\) with no accompanying charge current. The spatial separation of opposite spins towards opposite transverse edges creates a spin accumulation that can be detected electrically or optically. This mechanism is driven solely by the AM band anisotropy, without requiring spin-orbit coupling or external magnetic fields.

\subsection{Strain-Driven Net Spin Current and Piezomagnetism}
In the ideal lattice, the spin-dependent transport behavior is strictly constrained by the crystal symmetry, leading to a vanishing net spin polarization ~\cite{zhang_crystal-symmetry-paired_2025}. Overcoming this intrinsic compensation is a prerequisite for device integration, necessitating a strategic reduction of the system's symmetry. Consequently, we utilize uniaxial strain, which can effectively modulate the intrinsic spin states ~\cite{https://doi.org/10.1002/adma.201970322,https://doi.org/10.1002/smll.202402561}, to drive a structural phase transition that triggers valley polarization. Specifically, the $\overline{4}'$ magnetic symmetry, which maintains the energetic equivalence between the X and Y valleys in the pristine $\text{Ru}_2\text{MoSe}_4$, is effectively lifted under uniaxial strain along the $[100]$ or $[010]$ axis. As the lattice symmetry transitions from tetragonal to orthorhombic, the resulting symmetry breaking lifts the valley degeneracy and allows for the generation of controllable spin and valley currents.


The longitudinal spin-dependent conductivity $\sigma_{xx/yy}^{\sigma}$, where \(\sigma=\uparrow/\downarrow\) indicates spin-up/down, was investigated using the semi-classical Boltzmann transport equation under the relaxation time approximation. Here, $\sigma_{xx}^{\sigma}$ and $\sigma_{yy}^{\sigma}$ represent the longitudinal spin-dependent conductivities along the crystallographic $x$ and $y$ directions, respectively. While these two components are strictly identical in the pristine configuration due to the protection of the $\text{C}_4$ crystal symmetry, they exhibit pronounced divergence upon the application of uniaxial strain. This divergence ($\sigma_{xx}^{\sigma} \neq \sigma_{yy}^{\sigma}$) physically stems from the symmetry-breaking-induced structural anisotropy, which leads to distinct band dispersions along the $k_x$ and $k_y$ paths. Consequently, this generates direction-dependent group velocities ($v_{n\mathbf{k}, x} \neq v_{n\mathbf{k}, y}$), which are provided in SM, ultimately resulting in a highly anisotropic charge and spin transport behavior.

Uniaxial strain, serving as a critical symmetry-breaking mechanism, enables the generation of polarized currents whose direction is highly sensitive to the direction of uniaxial strain. We find that both the monolayer and AC-stacked bilayer $\text{Ru}_2\text{MoSe}_4$ exhibit similar behaviors. In the following, we focus our analysis on the AC-stacked bilayer configuration as a representative case, while the corresponding results for the monolayer $\text{Ru}_2\text{MoSe}_4$ are provided in Fig.~S3 in SM.

In the absence of SOC or external magnetic fields, the electronic states of AC-stacked bilayer $\text{Ru}_2\text{MoSe}_4$ at the X and Y valleys are degenerate, as shown in Fig.~\ref{fig:5}(b). Consequently, the conductivities of both channels are identical ($\sigma_{xx/yy}^{\uparrow} = \sigma_{xx/yy}^{\downarrow}$), yielding zero net spin polarization. However, as shown in Fig.~\ref{fig:5}(c-f), the application of strain breaks this symmetry. Under tensile strain ($6\%$), a significant valley polarization is induced specifically in the conduction band. The spin-down conductivity vanishes near 0.5 \(\text{eV}\), whereas the spin-up conductivity remains finite, creating a single-spin transport channel near the conduction band minimum (CBM).
Moreover, reversing the direction of the tensile strain can switch the spin polarization of the electronic states near the Fermi level. As the green shaded region in Fig.~\ref{fig:5}(d) and (f) indicate, this creates a regime where only one spin carrier is available for transport, leading to $\sigma_{xx}^{\downarrow} \gg \sigma_{xx}^{\uparrow}$ for $b$-strain and $\sigma_{yy}^{\uparrow} \gg \sigma_{yy}^{\downarrow}$ for $a$-strain, thereby enabling efficient electrical control of spin-polarized transport.
Furthermore, we observe a distinct difference in the response to compressive versus tensile strain. Under compressive strain, as shown in Fig.~\ref{fig:5}(c) and (e), valley polarization is induced in both the conduction and valence band. In contrast, tensile strain selectively induces valley polarization in the conduction band while the valence band edges remain nearly degenerate. 
This selective polarization demonstrates that tensile strain is a highly effective tool for engineering spin-filtering materials, as it concentrates the spin-splitting effect within a single carrier regime.


\begin{figure*}
    \centering
    \includegraphics[width=18cm]{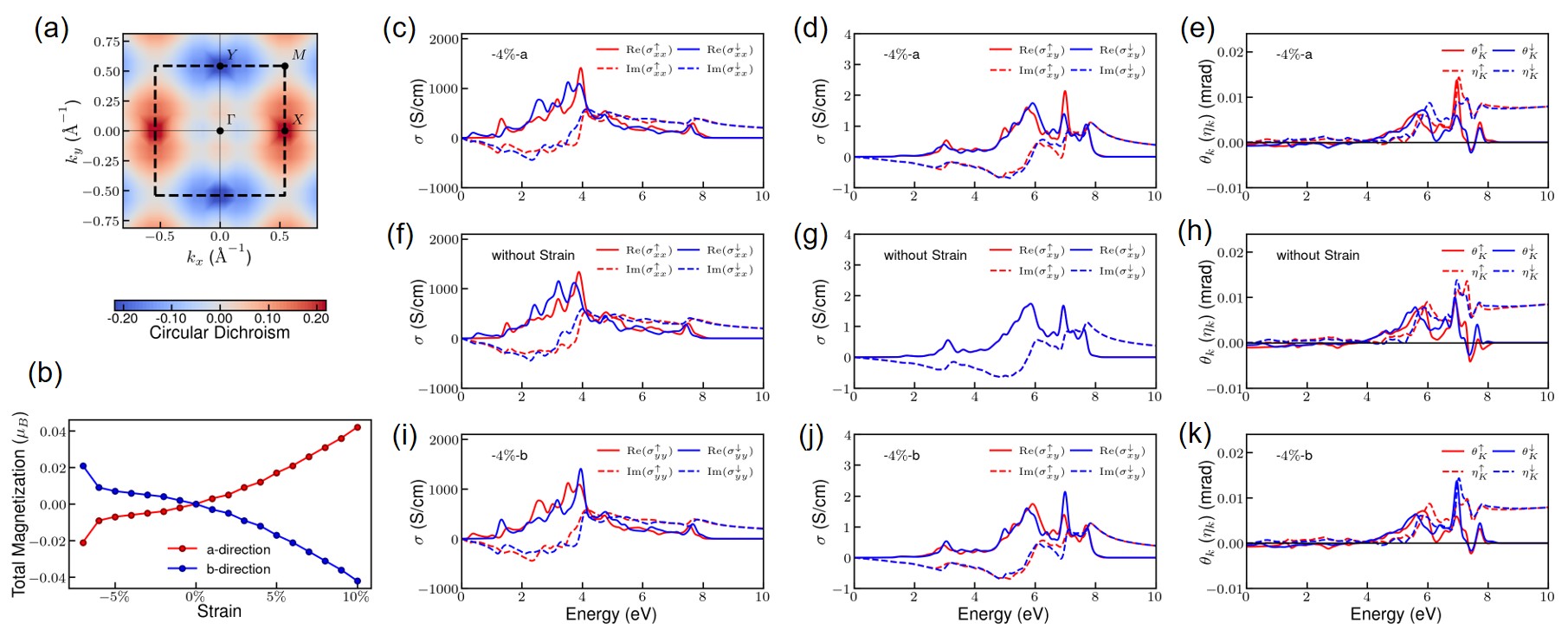}
    \caption{(a) CD spectra of monolayer $\text{Ru}_2\text{MoSe}_4$. (b) Net magnetic moment induced in monolayer $\text{Ru}_2\text{MoSe}_4$ as a function of uniaxial strain. (c–k) Calculated MOKE spectra under different strain conditions: (c–e) uniaxial compressive (\(4\%\)) strain along the $a$-direction, (f–h) the unstrained (pristine) case, and (i–k) uniaxial strain along the $b$-direction. The second, third, and fourth columns represent the diagonal optical conductivity \(\sigma_{xx/yy}\), off-diagonal optical conductivity \(\sigma_{xy}\), and Kerr signals (comprising both the Kerr rotation angle $\theta_K$ and ellipticity $\eta_K$), respectively.}
    \label{fig:3}
\end{figure*}

Beyond the generation of a net spin current, the anisotropy induced by uniaxial strain can further give rise to a net magnetization. 
As illustrated in Fig.~\ref{fig:3}(b), the evolution of the net magnetization under applied strain reveals a distinct piezomagnetic effect, where deformations along the $a$- and $b$-direction generate net magnetic moments with opposite signs. Besides, in AC-stacked bilayer $\text{Ru}_2\text{MoSe}_4$, the response of the net magnetic moment ($M_{total}$) to uniaxial strain also exists. As shown in Fig.~S8(b), the pristine system without strain effect exhibits zero total magnetization, which is a hallmark of its AM nature protected by the $S_{4z}\bm{T}$ symmetry. However, the application of uniaxial strain along either the $a$- or $b$-direction acts as a symmetry-breaking perturbation that lifts the exact compensation of the sublattices' magnetic moments. These phenomenon underscores a robust magnetic anisotropy that is inherently governed by the specific direction of lattice distortion, providing a magnetic signature for the symmetry-breaking process.
Crucially, uniaxial strain serves as a functional switch to activate and harvest usable net spin currents from the otherwise compensated AM's background, enabling a transition from a zero-net-spin state to a highly spin-polarized transport regime.

\subsection{Magneto-Optical Responses}
While the successful excitation and modulation of net spin currents have been demonstrated through electrical transport under uniaxial strain, the realization of highly integrated spintronic architectures necessitates non-invasive, non-contact techniques for information detection. Therefore, establishing an efficient optical readout is of paramount importance for the practical implementation of $\text{Ru}_2\text{MoSe}_4$-based devices. Motivated by this requirement, we further investigate the optical and MOKE of the system to evaluate its potential for multi-field spin-valley control.


As a hallmark of valleytronics, spin-valley locking dictates the valley-selective transport of carriers, which can be efficiently probed and manipulated via circular optical pumping. Our calculations, illustrated in Fig.~\ref{fig:3}(a), reveal that the X and Y valleys exhibit contrasting circular dichroism (CD) signatures, characterized by opposite signs of the Berry curvature-induced optical response. This divergence confirms the existence of valley-contrasting chiral selection rules, allowing for the selective excitation of carriers in a specific valley by tuning the helicity of the incident light. Consequently, this selective optical pumping not only generates significant valley polarization but also induces spin-polarized transport due to the robust coupling between valley and spin degrees of freedom. Such intrinsic CD underscores the fundamental role of broken time-reversal or inversion symmetry in defining valley physics, providing a viable optical route for generating valley- and spin-polarized currents even in the absence of external strain.

Building upon the static optical anisotropy revealed by CD, we further explore the dynamic MOKE of the system under external perturbations, which can be described as ~\cite{PhysRevB.92.144426,PhysRevB.96.214423}:
\begin{equation}
\theta_K + i\eta_K = -\frac{\sigma_{xy}}{\sigma_{xx/yy}\sqrt{1 + \frac{4\pi i}{\omega}\sigma_{xx/yy}}} 
\end{equation}
where \(\theta_k\) is the Kerr rotation angle, \(\eta_k\) is the Kerr ellipticity angle, $\sigma_{xy}$ and $\sigma_{xx/yy}$ are the off-diagonal and diagonal optical conductivity, and \(\omega\) is the frequency of incident light.
The application of uniaxial compressive strain serves as a decisive symmetry-breaking mechanism that lifts the spin degeneracy of $\sigma_{xy}$. As shown in Fig.~\ref{fig:3}(c-k), this results in a pronounced splitting between the spin-up and spin-down components of $\sigma_{xy}$ in both their real and imaginary parts.

In the monolayer configuration, a unique reciprocal correspondence is observed: the spin-up $\sigma_{xx}$ along the $a$-direction (both real and imaginary) quantitatively matches the spin-down $\sigma_{yy}$ along the $b$-direction. A similar mapping exists for the off-diagonal terms, where the spin-up $\sigma_{xy}$ under $a$-direction strain aligns with the spin-down $\sigma_{xy}$ under $b$-direction strain, which were shown in Fig.~\ref{fig:3}(d) and (j). This symmetry is further inherited by the MOKE signatures (Fig.~\ref{fig:3}(e) and (k)), with \(\theta_k\) and \(\eta_k\) exhibiting the same reciprocal characteristics. Such behavior provides a clear manifestation of CD in the anisotropic MOKE. These symmetry-driven effects persist in the bilayer system. Furthermore, the AC-stacked configuration significantly amplifies the magnitude of the MOKE signals compared to the monolayer counterpart, as shown in Fig.~S9. Specifically, the interlayer coupling and modified stacking symmetry in the AC bilayer lead to a dramatic enhancement of $\sigma_{xy}$, which increases by an order of magnitude from approximately $2~\text{S/cm}$ to $20~\text{S/cm}$.
This substantial enhancement of MOKE signals in the AC bilayer, combined with the aforementioned strain-tunable spin transport, establishes a robust foundation for the application of $\text{Ru}_2\text{MoSe}_4$ in the burgeoning field of spin-optical cross-disciplinary technologies.




\section{Conclusion}
In this study, we have comprehensively investigated the electronic, transport, and magneto-optical properties of the $\text{Ru}_2\text{MoSe}_4$ system by first-principles calculations and theoretical analysis, establishing it as a highly promising candidate for 2D AMs. 

Through applying in-plane uniaxial strain, we successfully break the exact magnetic compensatory symmetry inherent to the AM lattice. This strategic symmetry-breaking lifts the valley degeneracy, thereby activating a highly sensitive piezomagnetic response and generating robust, fully spin-polarized transport channels from a previously zero-net-spin background. Beyond semiclassical Boltzmann transport, we reveal that this structural modulation profoundly impacts the system's dynamic optical response. The transition to the AC-bilayer configuration drastically amplifies the off-diagonal optical conductivity, leading to an order-of-magnitude enhancement in the MOKE signal. This provides a highly efficient, non-contact optical readout mechanism for the underlying magnetic and valley states. 

Ultimately, the $\text{Ru}_2\text{MoSe}_4$ system serves as a versatile architecture where valley, spin, and layer degrees of freedom are intricately coupled. This work not only enriches the material reservoir for 2D AMs but also paves the way for future spintronic and magneto-optical devices capable of high-performance, non-volatile information processing through strain-mediated multi-field regulation.

\begin{acknowledgments}
This work was financially supported by the Fundamental Research Funds for the Central Universities (No.N25LPY025), the LiaoNing Revitalization Talents Program (Grant No. XLYC1907033) and the Natural Science Foundation of Liaoning province (Grant No.2023-MS-072). X. K. acknowledges the start up funding from Northeastern University, China.

There are no conflicts to declare.  
\end{acknowledgments}

\bibliographystyle{unsrt} 
\bibliography{references} 

\end{document}